\documentstyle[12pt]{article}
\newlength{\dinwidth}
\newlength{\dinmargin}
\begin{document}
{April 1995} \hfill{INP Report No. 1670/PH}
\vskip 2cm
\begin{center}
{\large \bf QCD Analysis of Deep Inelastic Diffractive Scattering}
\vskip 0.1cm
{\large \bf at HERA}
\vskip 0.5cm
{K. Golec--Biernat and J. Kwieci\'nski }
\vskip0.5cm
{\it Department of Theoretical Physics,
H. Niewodnicza\'nski Institute of Nuclear Physics,}

{\it ul. Radzikowskiego 152, 31-342 Krak\'ow, Poland}
\end{center}
\vskip1cm
\begin{abstract}
The QCD analysis of deep inelastic diffractive scattering is performed
assuming the dominance of the "soft" pomeron exchange and simple, physically
motivated parametrization of parton distributions in a pomeron. Results of
the analysis are compared with the recent data obtained by
the H1 collaboration at HERA.
Both the LO and NLO approximations are considered and the
theoretical predictions concerning the quantity $R=F_L^D/F_T^D$
for diffractive structure functions are presented.
\end{abstract}
\vskip1cm
The  diffractive deep inelastic scattering is the
process
\begin{equation}
e + p \rightarrow e^{\prime} + X + p^{\prime}
\label{largey}
\end{equation}
where there is a large
rapidity gap between the diffractively produced system $X$ and the recoiled
proton $p^{\prime}$ (or excited proton state).  To be precise, the process
(\ref{largey}) reflects the diffractive interaction of the virtual photon
(see Fig. 1a).  The recent measurements performed by  both ZEUS and H1
collaborations at the DESY ep collider HERA \cite{ZEUS,H11,H12} have shown
that the  diffractive structure function constitutes  a significant  part
  of the proton  structure function $F_2$.

The cross section for the process (1) is related in the following way to the
diffractive structure functions ${dF_2^D(x_P,\beta,Q^2,t)\over dx_P dt},
{dF_L^D(x_P,\beta,Q^2,t)\over dx_P dt}$:
\begin{displaymath}
{d\sigma^D \over dxdQ^2dx_Pdt} =
\end{displaymath}
\begin{equation}
{4\pi^2\alpha^2\over xQ^4} \left [ \left (1-y+{y^2\over2} \right )
{dF_2^D(x_P,\beta,Q^2,t)\over dx_P dt}-{y^2\over 2}
{dF_L^D(x_P,\beta,Q^2,t)\over dx_P dt} \right]
\label{sigmad}
\end{equation}
\newpage
{\noindent where the kinematical variables are defined below: }
\begin{eqnarray}
s=(p_e +p)^2 \approx 2p_ep~~~~~~    y={pq\over p_ep} \nonumber \\
Q^2=-q^2 ~~~~~~    x={Q^2\over 2pq} \nonumber \\
t=(p-p^{\prime})^2 ~~~    \beta={Q^2\over 2q(p-p^{\prime})} ~~~
x_P={x\over \beta}
\label{invar}
\end{eqnarray}
and where $q=p_e-p_e^{\prime}$.  The four momenta $p_e,p_e^{\prime},p$  and
$p^{\prime}$ correspond to those of the initial electron, final electron,
initial
proton and the recoiled proton respectively (see Fig. 1).\\

The main aim of this paper is to analyse the diffractive processes in deep
inelastic $ep$ scattering assuming that they are dominated by
a "soft" pomeron exchange
\cite{DOLA1}- \cite{COLLINS}.
Different  description of this processes  based on
the  "hard" pomeron which follows from perturbative QCD
 has been discussed in  \cite{KOLYA} -\cite{BARTELS2}.
The "hard" pomeron would however lead to much steeper
$x_P$ dependence than that  which is observed experimentally \cite{H12}.\\

Assuming that the diffractive process $\gamma ^{*}(Q^2)+p \rightarrow X +p$
is dominated by the pomeron exchange (see Fig. 1b)
with the pomeron being described as
a Regge pole with its trajectory
\begin{equation}
\alpha_P(t)=\alpha_P(0) + \alpha_P^{\prime}~t
\label{trajpom}
\end{equation}
where for the "soft" pomeron  $\alpha_P(0) \approx 0.08$, $\alpha_P^{\prime}
=0.25$ GeV$^{-2}$,
we get the following factorizable expression for the diffractive structure
functions :
\begin{equation}
{dF_{2,L}^D(x_P,\beta,Q^2,t)\over dx_P dt}=f(x_P,t)~F^P_{2,L}(\beta,Q^2,t)
\label{factor1}
\end{equation}
 The function $f(x_P,t)$ is the so called "pomeron flux factor" which,
 if the diffractively recoiled system is a single
proton,
has the following form \cite{COLLINS}:
\begin{equation}
f(x_P,t)=N~x_P^{1-2\alpha_P(t)}~ {B^2(t)\over 16 \pi}
\label{flux}
\end{equation}
where  $B(t)$ describes the pomeron coupling to a proton.  Its normalization
is such that the pomeron contribution to the pp total cross-section
$\sigma_{pp}(s)$ is
\begin{equation}
\sigma_{pp}(s)=B^2(0)\left({s\over s_0 }\right)^{\alpha_P(0)-1}
\end{equation}
with $s_0=1GeV^2$. The normalization factor $N$ was set to be equal to
${2 \over \pi}$ following the convention of refs. \cite{DOLA1,DOLA2,COLLINS}.
We have also set \cite{COLLINS}
\begin{equation}
B(t)= 4.6~mb^{1/2}~exp(1.9GeV^{-2}~t)
\label{bt}
\end{equation}
 and have slightly increased
$\alpha_P(0)$ in the flux factor  $f(x_P,t)$ (\ref{flux})
up to $\alpha_P(0)=1.1$
 \cite{H12}.  The slope of the pomeron trajectory
was kept equal to $\alpha_P^{\prime}=0.25$ GeV$^{-2}$.\\

The  functions $F^P_{2,L}(\beta, Q^2,t)$ are the pomeron structure
functions with the variable $\beta$ playing the role of the
Bjorken scaling variable for
the $\gamma^{*}(Q^2)$ pomeron inelastic "scattering". The factorization
property of the diffractive structure function into the pomeron
"flux" and the pomeron structure function is the direct consequence
of the single pomeron exchange mechanism of diffraction and of the assumption
that the pomeron is described by a single Regge pole.
 Factorization does not in general hold  in models
 which are based entirely on perturbative QCD
\cite{COLLINS,COLLSTRIK}.
Recent measurements \cite{H12}  have however found  that the
factorization of the diffractive structure functions agrees very well with
the  experimental data.
It has moreover been established that the $x_P$ dependence of the
flux factor  (integrated over $t$) can be parametrized as
$ x_P^{-n}$ with $n=1.19$.
This observation confirms dominance of the soft pomeron exchange
since the "hard" pomeron with its intercept $\alpha_P(0) \approx 1.5$
would give much steeper behaviour $\sim x_P^{-2}$ or so.\\

 In the region
of large $Q^2$ the pomeron structure function $F_{2}^P$
is expected to be described by the "hand-bag" diagram (see Fig. 2a)
\cite{DOLA1,DOLA2}.  It provides  the partonic description
of deep inelastic diffraction which
 leads to the Bjorken scaling mildly violated by the logarithmic
scaling violations implied by perturbative QCD.
In the QCD improved parton model the pomeron structure function
$F_{2}^P(\beta,Q^2,t)$
is related in  conventional
way to the quark $q_i^P(\beta,Q^2,t)$  distributions
in a pomeron:
\begin{equation}
F_{2}^P(\beta, Q^2,t) = 2~\beta~ \sum_i e_i^2~ q_i^P(\beta,Q^2,t)
\label{pmod}
\end{equation}
where $e_i$ are the quark charges (note that $q_i^P(\beta,Q^2,t)
= \bar q_i^P(\beta,Q^2,t)$).  The formula (\ref{pmod}) acquires
$O(\alpha_s)$ corrections in the next-to-leading approximation of
perturbative QCD. The next-to-leading approximation is also needed for
theoretically consistent introduction of the pomeron longitudinal
structure function
$F_{L}^P(\beta, Q^2,t)$.  In what follows we shall therefore discuss
both the leading and next to leading approximations.  In the leading
order approximation we shall set $F_L^P=0$.\\

At first we shall specify the details of the parton distributions in the
pomeron at the reference scale $Q^2=Q_0^2$ which we set equal to 4GeV$^2$.

At small $\beta$ both the quark and gluon distributions are assumed
to be dominated by the pomeron exchange as illustrated in the Fig. 2b.
\begin{eqnarray}
\beta~ q_i^P(\beta,Q_0^2,t)&=&a_i(t)~ \beta^{1-\alpha_P(0)} \nonumber \\
\beta~ g^P (\beta,Q_0^2,t)&=&a_g(t)~\beta^{1-\alpha_P(0)}
\label{smb}
\end{eqnarray}
The functions $a_i(t)$ and $a_g(t)$ can be estimated from the
factorization of
pomeron couplings \cite{DOLA1,BERGER,CAPELLA} (see Fig. 2b):
\begin{eqnarray}
a_i(t)=r(t)~a_i^p \nonumber \\
a_g(t)=r(t)~a_g^p
\label{factor2}
\end{eqnarray}
where the parameters $a_i^p$ and $a_g^p$
are the pomeron couplings
controlling the normalization of  the small $x$ behaviour of the
sea quark and gluon
distributions in a proton i.e.
\begin{eqnarray}
xq_i^p(x,Q_0^2)+x\bar q_i^p(x,Q_0^2)&=&2~a_i^p~x^{1-\alpha_P(0)} \nonumber \\
xg^p(x,Q_0^2)&=&a_g^p~x^{1-\alpha_P(0)}
\label{smxp}
\end{eqnarray}
and the function $r(t)$ is:
\begin{equation}
r(t)={\pi\over 2}{G_{PPP}(t)\over B(0)}
\label{rt}
\end{equation}
The coupling $G_{PPP}(t)$ is the triple pomeron coupling (see Fig.2 b) and
its magnitude can be estimated from the cross-section of the diffractive
production $p+\bar p \rightarrow p+X$ in the limit of large mass $M_X$
of the diffractively produced system $X$.  The relevant formula for this
cross-section is:
\begin{equation}
s{d\sigma \over dt dM_X^2} = {B^2(t)\over 16 \pi}~ G_{PPP}(t)~B(0)~
\left ({s\over M_X^2}\right )^{2\alpha_P(t)-1} \left ( {M_X^2\over s_0}
\right)^{\alpha_P(0)-1}
\label{difpp}
\end{equation}
(The factor ${\pi \over 2}$ in the eq. (\ref{rt}) follows from the convention
$N={2\over \pi}$ for the normalization factor $N$ in the eq. (\ref{flux})
\cite{DOLA1,DOLA2,COLLINS}).
We neglected the (weak) $t$ dependence of the function $r(t)$ and have
estimated its
magintude from the recent Tevatron data \cite{TEVAT} as
$r(t) \approx r(0)=0.089$.
The parameters $a_i^p$ were estimated assuming that the
sea quark distributions in a proton can be parametrized as:
\begin{equation}
 xq_i^p(x,Q_0^2)+x\bar q_i^p(x,Q_0^2)=2~a_i^p~x^{1-\alpha_P(0)}~(1-x)^7
\label{seap}
\end{equation}
and fixing the constants $a_i^p$ from the requirement that the average
momentum fraction which corresponds to those distributions is the same
as that which follows from the recent parametrization of parton distributions
in a proton \cite{MRSA,MRSG}.  The momentum sum rule has also been used
to fix the parameter $a_g^p$ i.e. we assumed
\begin{equation}
xg^p(x,Q_0^2)=a_g^p~x^{1-\alpha_P(0)}~(1-x)^5
\label{glup}
\end{equation}
and imposed the condition that the gluons carry 1/2 momentum of the proton.
We extrapolated the pomeron dominated  quark and gluon  distributions in a
pomeron
(see (\ref{smb})) to the
region of
arbitrary values of $\beta$ by multiplying the factor $\beta^{1-\alpha_P(0)}$
by $1-\beta$ \cite{BERGER}.

 We have also included the term
proportional to $\beta(1-\beta)$
in both the quark and gluon distributions \cite{BERGER}.  The normalization
of  this term in the quark distributions has been estimated in
\cite{DOLA2} assuming that it is dominated by the quark-box diagram
of Fig.2c with the non-perturbative couplings of pomeron to quarks.
In this model one gets:
\begin{equation}
\beta~ q^P(\beta,Q_0^2,t)={C \pi\over 3}~ \beta~(1-\beta)
\label{land}
\end{equation}
where $C \approx 0.17$ \cite{DOLA2}.
We found that the fairly reasonable description of data can be achieved
provided that the constant $C$  is enhanced
by a factor equal to 1.5.  We have also assumed that the relative
normalization of the quark distributions in a pomeron corresponding
to different flavours is the same as that of the sea quark distributions
in a proton \cite{MRSA,MRSG}.
Finally the normalization of the term proportional
to $\beta(1-\beta)$
in the gluon distribution in a pomeron has been obtained by imposing the
momentum sum rule.

As the result of the estimates
and extrapolations discussed above the parametrization of parton
distributions in a pomeron at the reference scale $Q_0^2=$ 4GeV$^2$
looks as follows:
\begin{eqnarray}
\beta~ u^P(\beta,Q_0^2,t)&=&0.4~(1-\delta)~ S^P(\beta) \nonumber \\
\beta~ d^P(\beta,Q_0^2,t)&=&0.4~ (1-\delta)~ S^P(\beta) \nonumber \\
\beta~ s^P(\beta,Q_0^2,t)&=&0.2~ (1-\delta)~ S^P(\beta) \nonumber \\
\beta~ c^P(\beta,Q_0^2,t)&=&\delta~ S^P(\beta) \nonumber \\
\beta~ g^P(\beta,Q_0^2,t)&=&(0.218~ \beta^{-0.08}~+~3.30~ \beta)~ (1-\beta)
\label{ppar}
\end{eqnarray}
where the function $S^P(\beta)$ is parametrized as below
\begin{equation}
S^P(\beta)= (0.0528~\beta^{-0.08}~+~0.801~\beta)~(1-\beta)
\label{spar}
\end{equation}
and $\delta$=0.02 \cite{MRSA,MRSG}. Following the approximations discussed
above we have neglected the $t$ dependence in those parton distributions.
The  analysis of the pomeron structure functions based
on different parametrizations of parton distributions in a pomeron has
recently been presented in refs. \cite{CAPELLA,STIRLING}.

The parton distributions defined by eqs.(\ref{ppar},\ref{spar}) were next
evolved up to the values of $Q^2$
for which the data exist using the LO Altarelli-Parisi evolution equations
\cite{AP,APR} with $\Lambda=0.255$ GeV \cite{MRSG}.
In Fig. 3a,b we show our results for the quantity:
\begin{equation}
 F_2^D(\beta,Q^2)=\int_{x_{PL}}^{x_{PH}}dx_P\int_{-\infty}^0 dt
{dF_{2}^D(x_P,\beta,Q^2,t)\over dx_P dt}
\label{dif}
\end{equation}
with $x_{PL}=0.0003$ and $x_{PH}=0.05$ \cite{H12}.  The theoretical curves
are compared with the recent data from the H1 collaboration
 at HERA \cite{H12}.
We find that the agreement with the data is very good concerning both the
shape in $\beta$ as well as the evolution with $Q^2$.
The following points should be noticed:
\begin{itemize}
\item
The diffractive structure function $F^D_2(\beta,Q^2)$at the moderate
values of $\beta$ is
dominated by "hard" component of the quark distributions in a pomeron (i.e.
that component which is proportional to $\beta(1-\beta)$).
Relatively small magnitude  of the soft component (i.e. of the term which
is proportional to $\beta^{-0.08}$) is a direct consequence of the small
magnitude
of the triple pomeron coupling (see the eqs. (\ref{smb}),(\ref{factor2}),
(\ref{ppar}) and (\ref{spar})).
\item
The soft component of the parton distributions is important at
the small values of $\beta \le 0.1$ or so. The agreement of our
parametrization with the data shows that the estimate of its normalization
 from the factorization of pomeron couplings is
quite reasonable. At large $Q^2$
the $\beta$ spectrum becomes softer   as  the result
 of the QCD evolution.
\item
It turns out that  gluons  carry more than 60 \% of the pomeron momentum.
The gluon content of  the pomeron can only indirectly manifest
itself through the QCD evolution of the diffractive structure function
with $Q^2$.
It should
be  observed at this point
that the relatively large gluon distributions in a pomeron
with a "hard" $1-\beta$
spectrum leads
to the increase of the diffractive structure functions with increasing $Q^2$
for fixed $\beta$ , up to
rather relatively large values of $\beta \approx 0.5$.  It may be seen
from Fig. 3a that this effect is nicely confirmed by  the data.
It should be noted at this point that  the proton
structure function $F_2^p(x,Q^2)$
starts to decrease with increasing $Q^2$ already for $x \ge 0.1$.

\end{itemize}

We have also performed the NLO QCD analysis in order to be able to
consistently
introduce the quantity $R=F_L^D(\beta,Q^2)/F_T^D(\beta,Q^2)$ where
$F_T^D(\beta,Q^2)=F_2^D(\beta,Q^2)-F_L^D(\beta,Q^2)$ and
\begin{equation}
F_L^D(\beta,Q^2)=\int_{x_{PL}}^{x_{PH}}dx_P\int_{-\infty}^0 dt
{dF_{L}^D(x_P,\beta,Q^2,t)\over dx_P dt}
\label{fl}
\end{equation}
We compare the NLO and LO results for $F_2^D(\beta,Q^2)$ in Fig. 4 a,b.
We can see that both approximations lead to similar
predictions.In Fig.4 c,d we show our
results for $R$.
The longitudinal
diffractive structure function is mainly driven by the gluon distributions
in a pomeron
and a large amount of gluons (see \ref{ppar})
implies that $R$ can reach
 $0.5$ for $\beta \le 0.1$.
It may also be seen from  Figs. 4 a,b  that $R$ exhibits  much more
softer dependence on $\beta$  than the structure function $F_2^P(\beta,Q^2)$.
\\

To summarize,  we have presented in this paper the QCD analysis of the
diffractive structure function within the simple soft pomeron exchange
picture.
In this model the diffractive structure functions factorize into the
flux factor and the pomeron structure functions which are sensitive on
the parton
content of the pomeron. We based our analysis on the simple parametrization
of the parton distributions which utilized both the factorization
of pomeron couplings and the quark-box contribution and obtained a good
description of the experimental data.  We have also performed the
NLO QCD analysis and estimated the magnitude of $R
=F_L^D(\beta,Q^2)/F_T^D(\beta,Q^2)$ (see Fig.4 c,d).  Its knowledge
should be useful for  more refined experimental determination of the
diffractive structure function $F_2^D$.\\

{\bf Acknowledgments}
\par
We thank Albert de Roeck for several very instructive discussions which
prompted this study. This research has been supported in part by the
Polish State Committee for Scientific Research grants N0s 2 P302 062 04
and 2 P03B 231 08.

\newpage
{\Large {\bf Figure Captions}}
\begin{enumerate}
\item
(a)Kinematics of the large rapidity gap process $e+p \rightarrow e^{\prime}
+X+p^{\prime}$.The wavy line represents the virtual photon. \\
(b)The Pomeron exchange diagram for diffractive production of the
hadronic system $X$ by a virtual photon.  The wavy and zigzag lines
represent the virtual photon and pomeron respectively.
\item
(a)The hand bag diagram for the virtual Compton diffractive production.
The wavy and zigzag lines represent the virtual photon and pomeron
respectively. The continuous lines in the upper part of the diagram
denote quarks (antiquarks).\\
(b)The triple pomeron diagram contribution to the "hand-bag" diagram
of Fig. 2a.  The coupling $G_{PPP}$ denotes the triple pomeron coupling.\\
(c)The quark box diagram contribution to the "hand-bag" diagram of Fig. 2a.
\item
Theoretical predictions for the
 diffractive structure function $F_2^D(\beta, Q^2)$
 defined by the formula (\ref{dif}) and their comparison with the data
from HERA \cite{H12}.  The structure function $F_2^D(\beta, Q^2)$ is
plotted  (a) as the function of $Q^2$ for fixed values of $\beta$  and
(b) as the function of $\beta$ for fixed values of $Q^2$.
\item
(a) Comparison of LO (continuous lines) with NLO (dashed lines)
results for the structure function
$F_2^D(\beta, Q^2)$ defined by the eq. (\ref{dif}) plotted  as
the function of $Q^2$ for fixed values of $\beta$.\\
(b) Comparison of LO (continuous lines) with NLO (dashed lines)
results for the structure function
$F_2^D(\beta, Q^2)$ defined by the eq. (\ref{dif}) plotted  as
the function of $\beta$ for fixed values of $Q^2$.
\item
(a)Theoretical predictions for the quantity $R={F_L^D\over F_T^D}$
  plotted  as
the function of $Q^2$ for fixed values of $\beta$.\\
(b)Theoretical predictions for the quantity $R={F_L^D\over F_T^D}$
  plotted  as
the function of $\beta$ for fixed values of $Q^2$.
\end{enumerate}
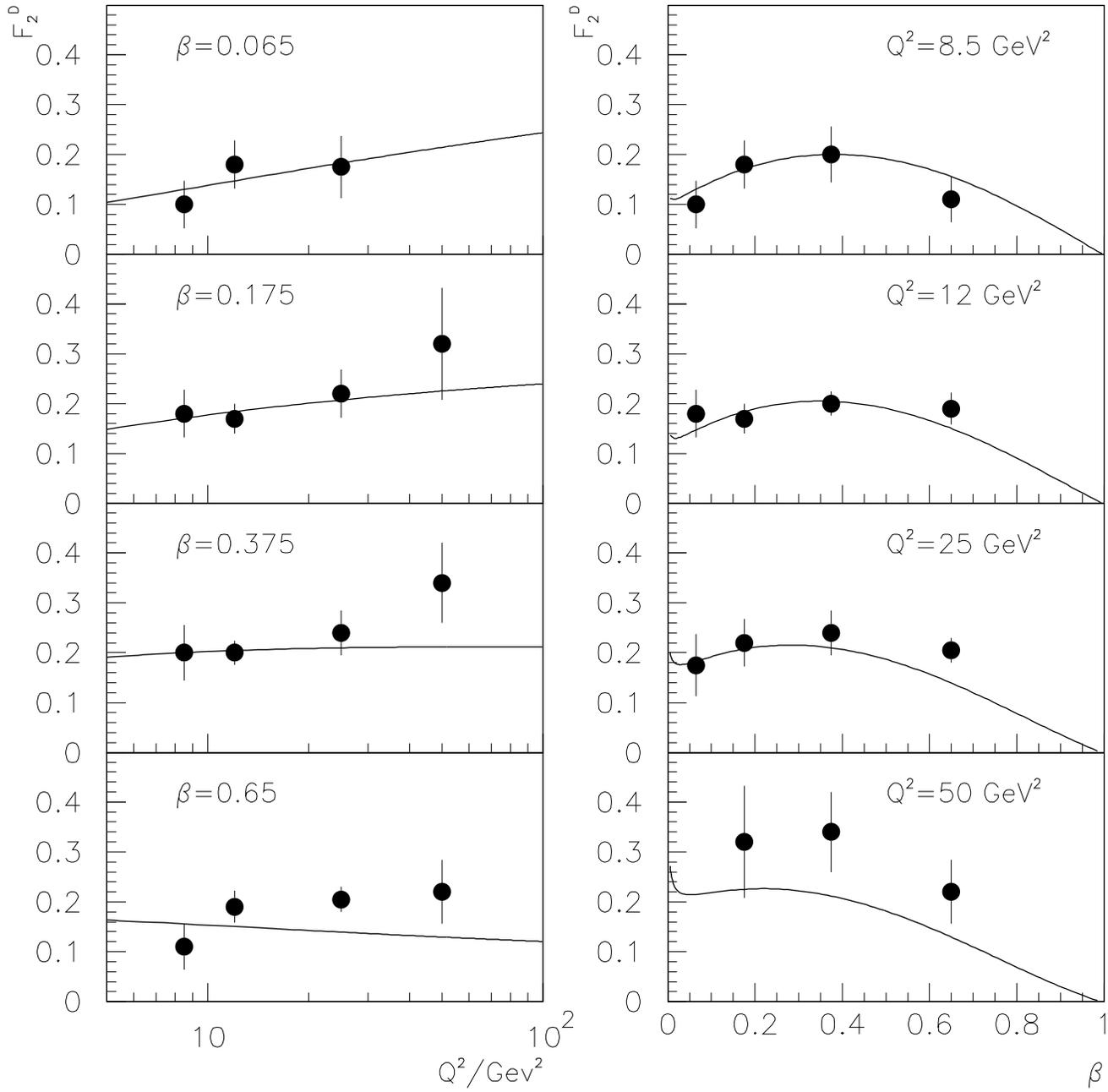
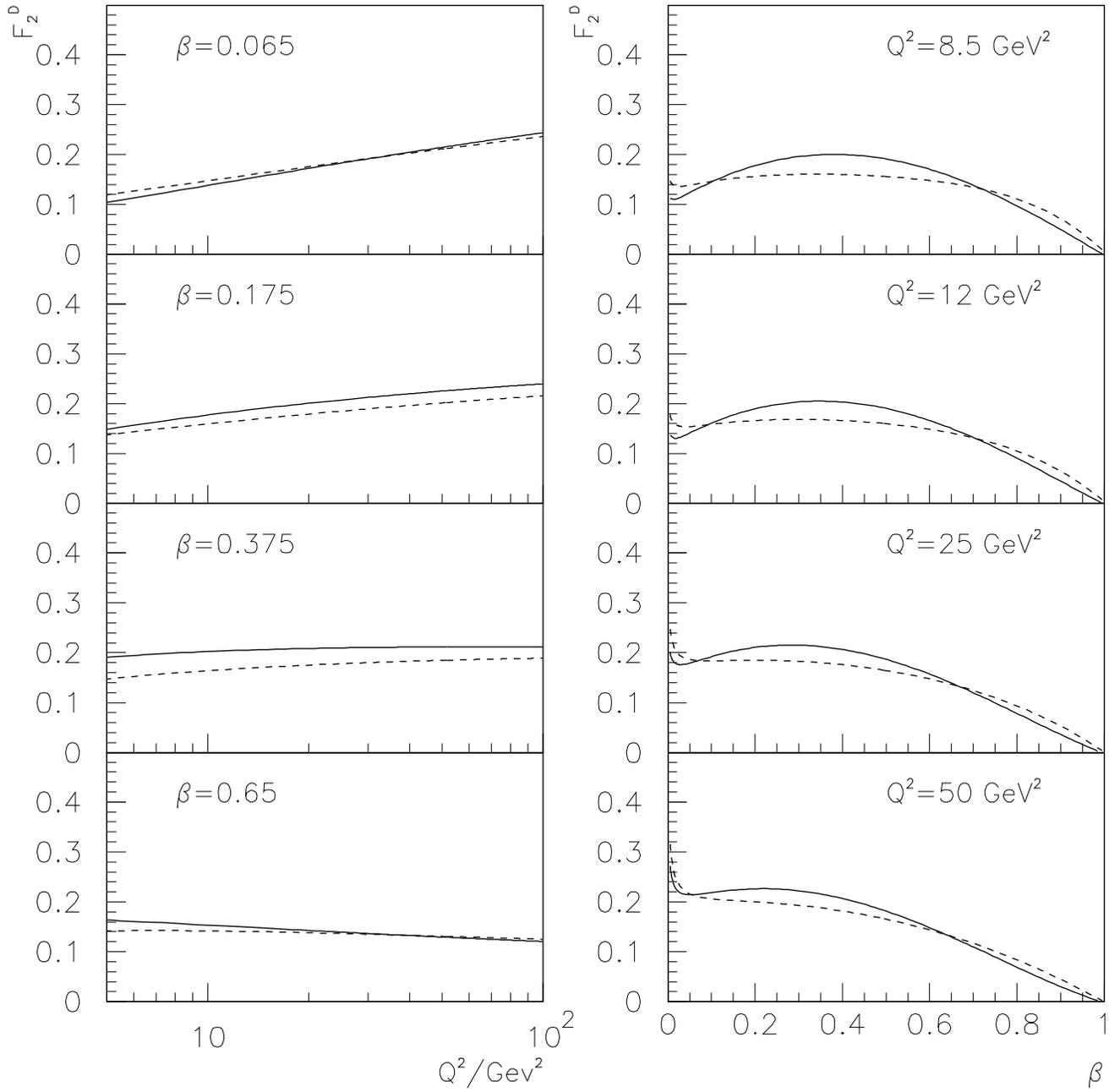
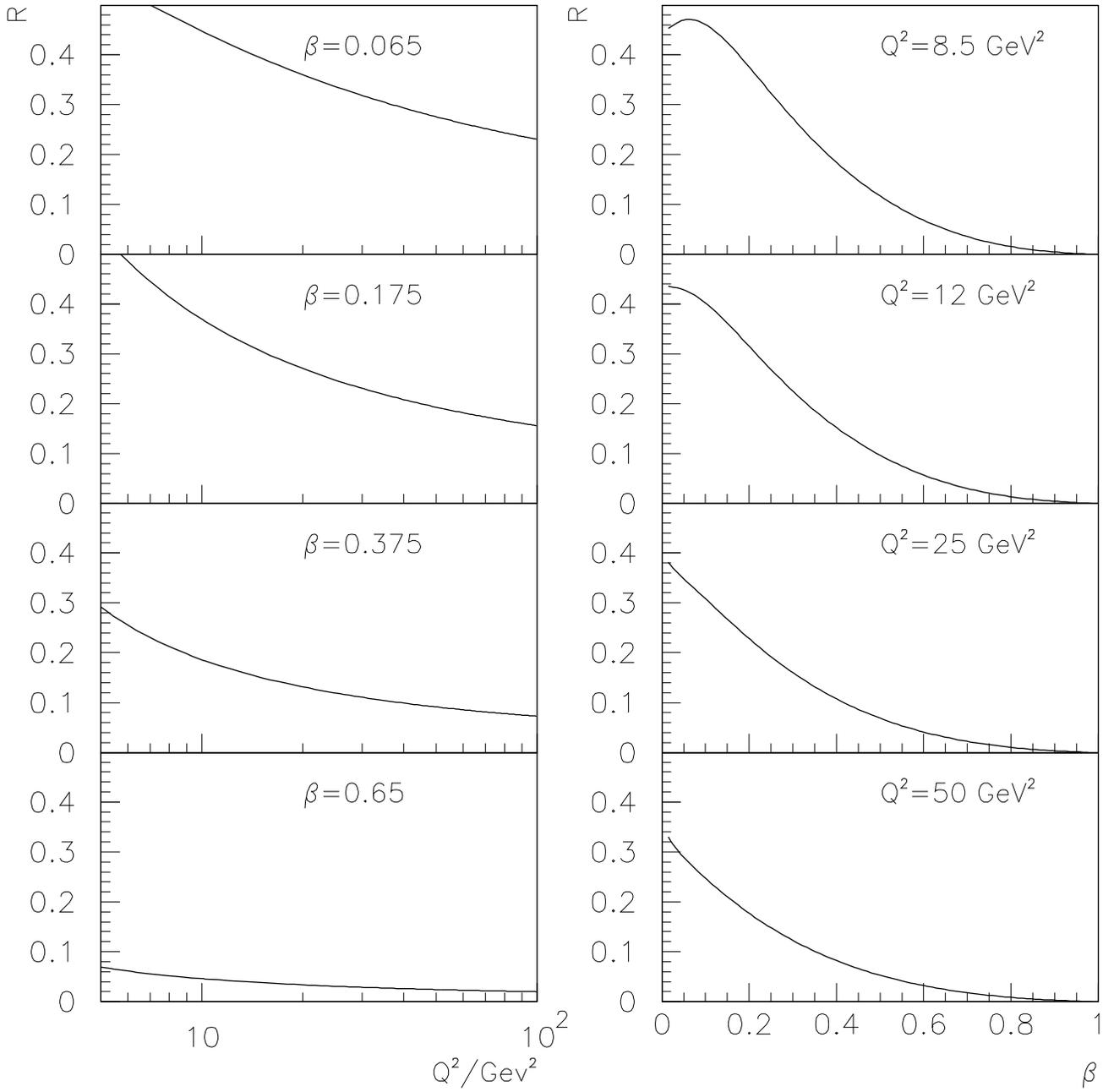
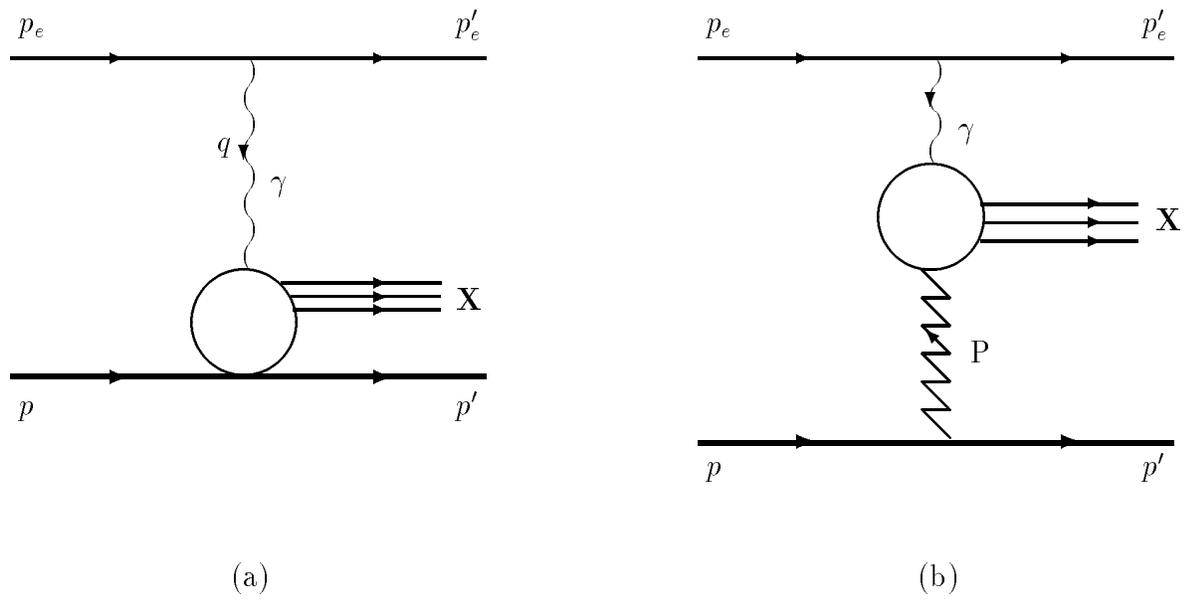
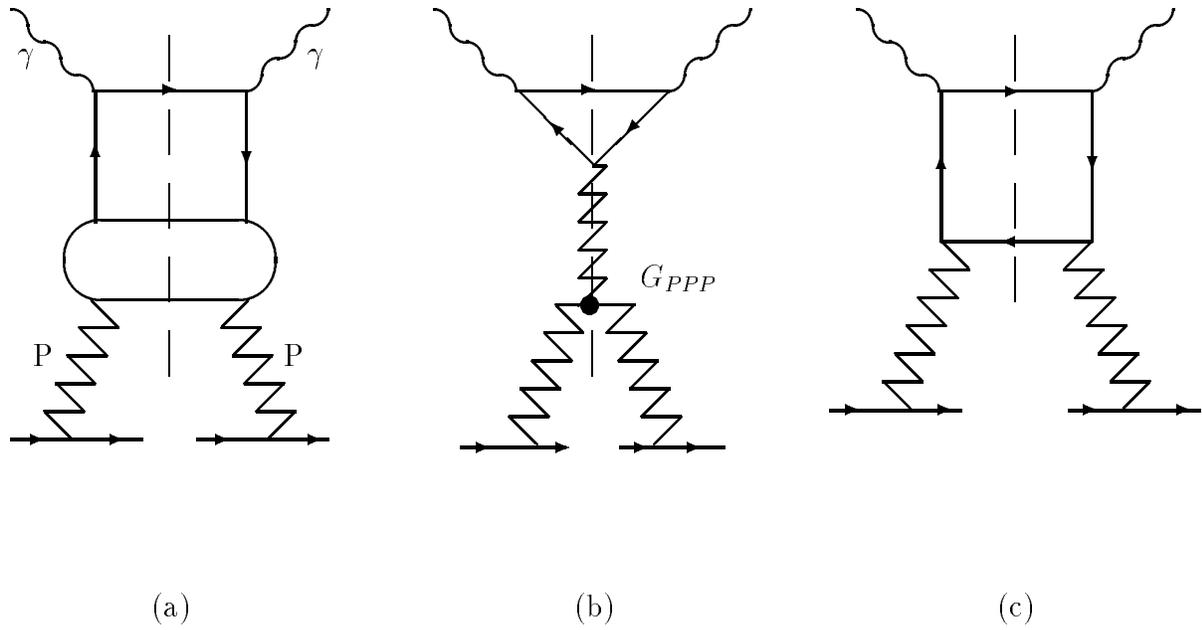
\end{document}